\documentclass[aip,jap,reprint,groupedaddress,showpacs]{revtex4-1}

\usepackage{graphicx}
\usepackage{dcolumn}
\usepackage{bm}

\begin{document}

\title{Low-temperature thermal conductivity of antiferromagnetic
$S$ = 1/2 chain material CuCl$_2$$\cdot$2((CH$_3$)$_2$SO)}

\author{W. P. Ke$^{1}$}

\author{J. Shi$^{2}$}

\author{F. B. Zhang$^{1}$}

\author{Z. Y. Zhao$^{1}$}

\author{C. Fan$^{1}$}

\author{X. Zhao$^{3}$}
\email{xiazhao@ustc.edu.cn}

\author{X. F. Sun$^{1}$}
\email{xfsun@ustc.edu.cn}

\affiliation{$^{1}$Hefei National Laboratory for Physical Sciences
at Microscale, University of Science and Technology of China,
Hefei, Anhui 230026, People's Republic of China}

\affiliation{$^{2}$Department of Physics, University of Science
and Technology of China, Hefei, Anhui 230026, People's Republic of
China}

\affiliation{$^{3}$School of Physical Sciences, University of
Science and Technology of China, Hefei, Anhui 230026, People's
Republic of China}


\begin{abstract}

We study the heat transport of $S = 1/2$ chain compound
CuCl$_2$$\cdot$2((CH$_3$)$_2$SO) along the $b$ axis (vertical to
the chain direction) at very low temperatures. The zero-field
thermal conductivity ($\kappa$) shows a distinct kink at about 0.9
K, which is related to the long-range antiferromagnetic (AF)
transition. With applying magnetic field along the $c$ axis,
$\kappa(H)$ curves also show distinct changes at the phase
boundaries between the AF and the high-field disordered states.
These results indicate a strong spin-phonon interaction and the
magnetic excitations play a role in the $b$-axis heat transport as
phonon scatterers.

\end{abstract}


\maketitle

Low-dimensional quantum magnets were revealed to exhibit exotic
ground states, magnetic excitations, and quantum phase transitions
(QPTs).\cite{Sachdev} The heat transport in these materials has
attracted much attention due to the role of magnetic
excitations.\cite{Heidrich, Hess, Sologubenko1} In quasi-one
dimensional systems, particularly for the $S$ = 1/2 chains or spin
ladders, a large contribution of magnetic excitations to
transporting heat along the spin chains was theoretically
predicted and experimentally confirmed.\cite{Heidrich, Hess,
Sologubenko1} Along other directions, however, the magnetic
excitations are weakly dispersive and they cannot carry heat but
scatter phonons. In some examples, such spin-phonon interactions
are so strong that the thermal conductivity ($\kappa$) show
remarkable feature of resonant phonon scattering and can be
changed by magnetic field very strongly.\cite{Sologubenko2,
Sun_DTN, Chen_MCCL, Zhao_BCVO}

CuCl$_2$$\cdot$2((CH$_3$)$_2$SO) (abbreviated as CDC) is an $S =
1/2$ spin chain material.\cite{Kenzelmann1, Kenzelmann2, Chen} It
crystallizes in an orthorhombic structure, with the space group
$Pnma$ and the room-temperature lattice constants $a =$ 8.054, $b
=$ 11.546, and $c =$ 11.367 \AA.\cite{Willett} The Cu-Cl-Cu bonds
form spin chains along the $a$ axis with strong superexchange
interactions between Cu$^{2+}$ spins mediated by Cl$^-$
ions.\cite{Kenzelmann2, Chen} The magnetic structure was
determined as an AF chain system with the intrachain interactions
$J = 1.43$ meV.\cite{Landee1, Landee2, Kamieniarz} Inelastic
neutron scattering revealed that the spectra of magnetic
excitations include both the soliton and the bound-spinon states
in the presence of staggered fields.\cite{Kenzelmann1,
Kenzelmann2} The specific heat results showed a long-range AF
transition at $T_N = 0.93$ K, which changes with applying magnetic
fields.\cite{Chen} When the field is along the $b$ axis, the AF
state is enhanced and the transition temperature increases with
increasing field.\cite{Chen} With magnetic fields along the $a$ or
the $c$ axis, the transition moves to lower temperatures with
increasing field and disappears at about 6 and 3.9 T for $H
\parallel a$ and $c$, respectively, which corresponds to a
second-order QPT from the magnetically ordered state to a disorder
state.\cite{Chen} In addition, a spin-flop transition occurs at
about 0.3 T ($H \parallel c$), due to the competitions among the
spin anisotropy, the interchain interactions, and the staggered
fields.\cite{Kenzelmann1, Kenzelmann2, Chen} CDC was recently
found to exhibit magnetically induced multifferroicity and
therefore the spin-phonon coupling in this material would be an
interesting physical issue.\cite{Zapf}

Here we report a study on the thermal conductivity ($\kappa$) of
CuCl$_2$$\cdot$2((CH$_3$)$_2$SO) single crystal at very low
temperatures down to 0.3 K and in magnetic fields up to 14 T. It
is found that, for the heat current vertical to the spin-chain
direction, the magnetic excitations play a role of scattering
phonons. Applying magnetic field along the $c$ axis can suppress
the magnetic excitations and lead to a large increase of $\kappa$
at high fields.

The CDC single crystals with nearly a parallelepiped shape were
grown using a solution method. The largest naturally formed
surface is the $ab$ plane and the longest edge is along the $b$
axis (the spin chains are along the $a$ axis). The thermal
conductivities were measured along the $b$ axis by using a
conventional steady-state technique.\cite{Sun_DTN, Chen_MCCL,
Zhao_BCVO} The specific heat was measured by the relaxation method
in the temperature range from 0.4 to 30 K using a commercial
physical property measurement system (PPMS, Quantum Design). The
magnetic susceptibility and magnetization were measured using a
SQUID-VSM (Quantum Design). One difficulty in this work is that
the CDC crystals are so fragile that they are easily broken upon
cooling. It is almost impossible to carry out the low-$T$ $\kappa$
measurements on a CDC crystal without introducing any damage. In
fact, every time after a thermal cycle, the crystals were broken
into several pieces or some obvious cracks appeared. In this work,
the thermal conductivity data were collected on a sample with some
cooling-produced cracks. Therefore, it should noted that the
magnitude of $\kappa$ is likely smaller than the intrinsic value.
However, the qualitative behaviors of $\kappa$, particularly the
field dependencies, are able to demonstrate the role of magnetic
excitations in the heat transport properties.

\begin{figure}
\includegraphics[clip,width=5.0cm]{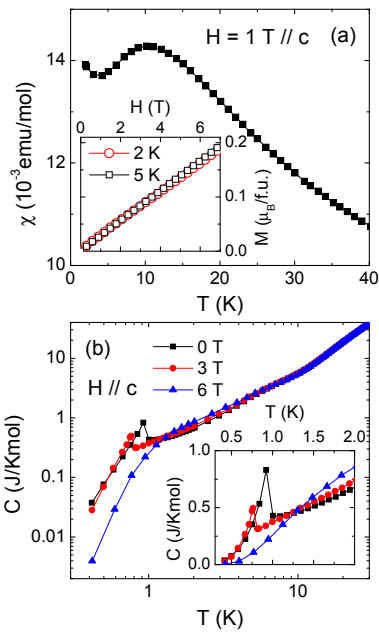}
\caption{(Color online) (a) Magnetic susceptibility of a CDC
single crystal measured in 1 T field along the $c$ axis. The inset
shows the magnetization curves measured with $H \parallel c$ and
at 2 and 5 K. (b) Temperature dependencies of specific heat of a
CDC single crystal from 0.4 to 30 K and with $H \parallel c$. The
inset shows the low-temperature data in linear plot.}
\end{figure}

Figure 1(a) shows the low-$T$ magnetic susceptibility and
magnetization of a CDC single crystal with $H \parallel c$. A
broad peak of $\chi(T)$ at about 10 K demonstrates the
low-dimensional characteristic of the spin systems. The
magnetization shows a good linear behavior at temperatures down to
2 K. These results are essentially consistent with the earlier
results.\cite{Landee1, Landee2}

Figure 1(b) shows the low-$T$ specific heat data of a CDC single
crystal with $H \parallel c$. In zero field, the curve displays an
obvious deviation from the $T^3$ law (the simple crystal lattice
specific heat) below 12.5 K, which is caused by the enhancement of
the magnetic correlations. At lower temperatures, a sharp peak
shows up at 0.92 K, which is known to be a long-range AF
transition. Applying magnetic field along the $c$ axis can
suppress the AF order with the transitions peak shifting to lower
temperatures. Note that the data show exponential $T$-dependence
at very low temperatures, indicating a gapped magnetic excitation.
These phenomena are consistent well with the earlier
results.\cite{Chen} An interesting phenomenon is that in high
field (6 T), while the AF transition peak disappears and the
low-$T$ specific heat is strongly suppressed, the specific heat at
high temperatures is somewhat enhanced. As a result, the $C(T)$
curve shows a pronounced hump at low temperatures. Apparently, the
spin fluctuations are still very strong in high magnetic fields.

\begin{figure}
\includegraphics[clip,width=5.0cm]{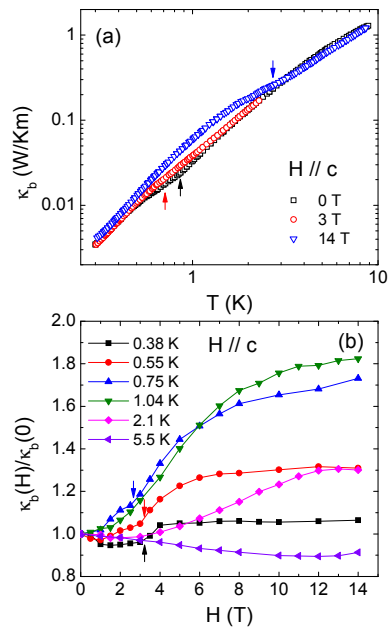}
\caption{(Color online) Temperature dependencies (a) and
magnetic-field dependencies (b) of thermal conductivity of a CDC
single crystal for $H \parallel c$ and heat current $J \parallel
b$. The sample size is $4.05 \times 1.04 \times 0.55$ mm$^3$. The
arrows indicate the distinct features in these curves, as
discussed in the main text.}
\end{figure}

Figure 2(a) shows the temperature dependencies of $\kappa$ with $H
\parallel c$ and heat current $J \parallel b$. At zero field,
$\kappa(T)$ displays a kink-like feature at $T <$ 1 K, which is
apparently related to the AF transition shown by the specific heat
data. In principle, this feature can be caused by either a magnon
heat transport below $T_N$ or a strong phonon scattering by
magnetic excitations at the critical region of phase transition.
In the present work, the $\kappa$ was measured along the $b$ axis,
in which the spin interactions are weak. Therefore, the magnetic
excitations can influence the heat transport only by scattering
phonons; that is, the feature of the zero-field $\kappa(T)$ at $T
<$ 1 K is likely related to a phonon scattering by the critical
fluctuations. Upon applying magnetic field to 3 T, the kink-like
feature becomes weaker and shifts to lower temperature, which
corresponds to the suppression of the AF transition. In a high
magnetic field of 14 T, this feature disappears while another
kink-like feature shows up at about 2.5 K, which is likely due to
the phonon-resonance scattering induced by an energy gap of the
spin spectrum. This field-induced gap is related to the presence
of a staggered $g$-tensor and Dzyaloshinskii-Moriya
interactions.\cite{Chen} Using either the boson model or the
sin-Gordon model, the gap was estimated to be about 0.8 meV at 14
T field along the $c$ axis.\cite{Chen} In a scenario of resonant
phonon scattering by magnetic excitations associated with a spin
gap, the strongest scattering occurs when the thermal energy
$k_BT$ and the gap size $\Delta$ has a relationship of $\Delta =
3.8 k_BT$.\cite{Berman, Sun_GBCO} Thus, the phonon resonant
scattering is expected to locate at about 2.4 K, which matches
well with the experimental data shown in Fig. 2(a) (an arrow
indicates the kink of 14 T data at $\sim$ 2.7 K).

In passing, it is worthy of pointing out that the thermal
conductivity along the spin-chain direction (the $a$ axis) is
expected to display the magnon heat transport, because of the
significant magnon dispersion along the chain. However, the
measurement along the $a$ axis was not yet successful for CDC
crystals, because one needs to cut the crystal along the direction
vertical to the naturally formed longest dimension and would very
easily damage this fragile material.

Figure 2(b) shows the detailed field dependencies of $\kappa$ with
$H \parallel c$ and $J \parallel b$. At lowest temperature (0.38
K), the $\kappa(H)$ isotherms show a weak decrease at low fields
and a small but sharp increase at $\sim$ 3.25 T; above 4 T , the
$\kappa$ is nearly independent of field. At higher temperatures
(0.55 -- 1.04 K), the $\kappa(H)$ show similar behavior but with
some small differences. First, the enhancement of $\kappa$ in high
fields is more significant. Second, the increase of $\kappa$ at
$\sim$ 3.25 T becomes not so sharp and moves to lower fields. The
high-field plateau is formed at much higher field or not formed up
to 14 T. For even higher temperature of 2.1 K, the field
dependence becomes weaker again and shows a smooth increase
function. At 5.5 K, the field dependence changes sign; that is,
the $\kappa$ is suppressed with applying field although it seems
to be recovering at very high fields. In general, the low-field
suppression of $\kappa$ is caused by the magnetic excitations
scattering phonons, while the high-field increase of $\kappa$ is
due to the suppression of magnetic excitations. Note that the
sharp increases of the low-$T$ $\kappa(H)$ curves have a good
correspondence to the transition from the low-field AF state to
the high-field paramagnetic state, which has already been
determined in some earlier works by using other
measurements,\cite{Chen, Zapf} as shown in Fig. 3.

\begin{figure}
\includegraphics[clip,width=6.0cm]{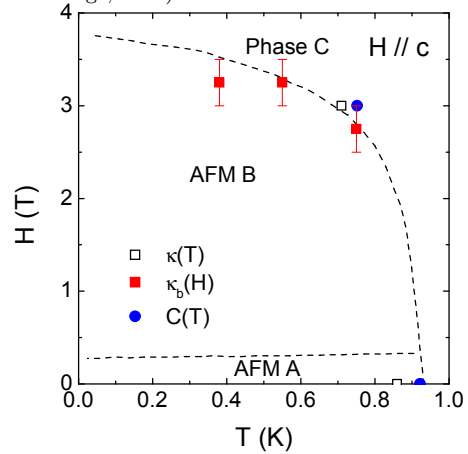}
\caption{(Color online) $H-T$ phase diagram of CDC for $H
\parallel c$. The dashed lines show the phase boundaries from
Refs. \onlinecite{Chen} and \onlinecite{Zapf}. AFM represents the
low-$T$ antiferromagnetic state while A and B phases are separated
by a spin-flop transition. Phase C is the paramagnetic state.}
\end{figure}

It is common to many magnetic materials that the magnetic
excitations can interact with phonons and sometimes introduce very
strong magnetic-field dependence of thermal
conductivity.\cite{Sologubenko2, Sun_DTN, Chen_MCCL, Zhao_BCVO,
Wang_HMO, Zhao_GFO} In particular, all the multiferroic materials,
of which the heat transport properties have been
studied,\cite{Wang_HMO, Zhao_GFO} are known to exhibits strong
interactions between magnetic excitations and phonons. Our data
show that the field dependence of $\kappa$ in CDC is comparable to
many other materials, which indicates a rather strong spin-phonon
coupling in this organic material.

We thank W. Tao for growing some of the CDC single crystals. This
work was supported by the National Natural Science Foundation of
China, the National Basic Research Program of China (Grants No.
2009CB929502 and No. 2011CBA00111), and the Fundamental Research
Funds for the Central Universities (Program No. WK2340000035).

\end{document}